# Optical data transmission field trial @ 44 Tb/s with a 49GHz Kerr soliton crystal microcomb

Mengxi Tan[1], Xingyuan Xu[1,2], and David J. Moss[1], *Fellow IEEE*
[1]Optical Sciences Centre, Swinburne University of Technology, Hawthorn, VIC 3122, Australia
[2]Dept. of Electrical and Computer Systems Engineering, Monash University, Clayton, 3800 VIC, Australia

*Abstract*—We report world record high data transmission over standard optical fiber from a single optical source. We achieve a line rate of 44.2 Terabits per second (Tb/s) employing only the C-band at 1550nm, resulting in a spectral efficiency of 10.4 bits/s/Hz. We use a new and powerful class of micro-comb called soliton crystals that exhibit robust operation and stable generation as well as a high intrinsic efficiency that, together with an extremely low spacing of 48.9 GHz enables a very high coherent data modulation format of 64 QAM. We achieve error free transmission across 75 km of standard optical fiber in the lab and over a field trial with a metropolitan optical fiber network. This work demonstrates the ability of optical micro-combs to exceed other approaches in performance for the most demanding practical optical communications applications.

## I. Introduction

Kerr micro-combs [1-4] offer the full potential of their bulk counterparts [5,6] but in an integrated footprint, since they generate optical frequency combs in integrated micro-cavity resonators. The realization of soliton temporal states called dissipative Kerr solitons (DKSs) [7-11] opened up a new method of mode-locking micro-combs that has in turn underpinned major breakthroughs in many fields such as spectroscopy [12,13], microwave and RF photonics [14], optical frequency synthesis [15], optical ranging including LIDAR [16, 17], quantum photonic sources [18-21], metrology [22, 23] and much more. One of their most promising applications has been in the area of optical fibre data communications where they have formed the basis of massively parallel multiplexed ultrahigh capacity optical data transmission [4, 24 - 26]. In this paper, [26] by employing a powerful new type of micro-comb based on soliton crystals [11], we report a world record speed of data transmission across standard optical fibre from any single optical source. We achieve a line rate of 44.2 Terabits/s (Tb/s) utilizing only the telecom 1550nm C-band, and achieve a very high spectral efficiency of 10.4 bits/s/Hz. Spectral efficiency is a critically important performance parameter since it directly governs how much total bandwidth can be realized in a system. Soliton crystals display very stable and robust operation and generation as well as a very high intrinsic conversion efficiency that, all taken together with the extremely low soliton micro-comb FSR spacing of 48.9 GHz that we achieve, enabled us to use a record high modulation coherent data format of 64 quadrature amplitude modulation (QAM). We demonstrate error free data transmission across a 75 km distance of standard optical fibre in our lab, but more importantly in a real-world field trial in an installed metropolitan area optical fibre testbed network in the Melbourne region. Our results were significantly helped by the capacity of the soliton crystals to work without any stabilization or feedback control at all, but only with very simple open loop systems. This significantly reduced the amount and sophistication of the instrumentation required. Our work directly proves the capability of optical Kerr microcombs to out-perform any other approach for practical demanding optical communications systems.

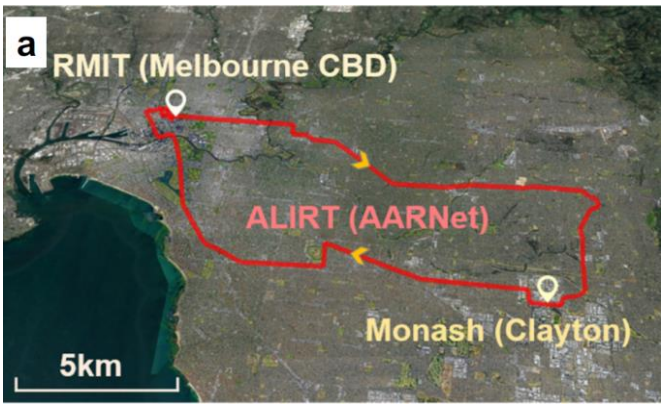

Figure 1. Field trial network in greater Metropolitan Melbourne.

Currently, 100's of Terabits/s are transmitted every instant across the world's fibre optic networks and the global bandwidth is growing at a rate of 25% /yr [27]. Ultrahigh capacity data links that use parallel massive wavelength division multiplexing (WDM) systems combined with coherent advanced modulation formats [28], are critical to meet this demand. Space-division multiplexing (SDM) is another emerging approach where multiple signals are transmitted either over multiple core or multiple mode fibre, or both [29]. In parallel with all of this, there is a growing movement towards very short links but still with very high capacity, particularly for data centres. Even just ten years ago, long haul networks such as undersea links spanning thousands of kilometres, used to dominate the global infrastructure, but nowadays the demand has dramatically shifted towards smaller scale applications including the aforementioned data centres as well as metropolitan area networks (tens to hundreds of kilometres in size). These trends demand highly compact, energy efficient and low-cost devices. Photonic integrated circuits are the only approach that can address these needs, where the optical source is absolutely key to each link, and therefore has the greatest need to meet these requirements. The capability of generated all of the wavelengths by a single chip that is both integrated and compact in order to replace multiple lasers, will yield the highest benefits [30-32].

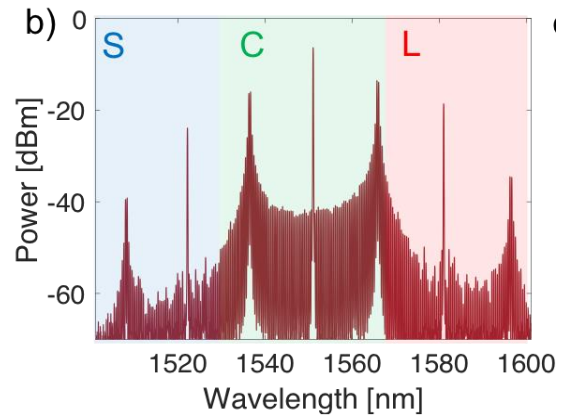

Figure 2. Soliton crystal spectrum.

Oscillation states in micro-resonators that have a crystalline type of profile along the resonator path, forming in the angular domain of

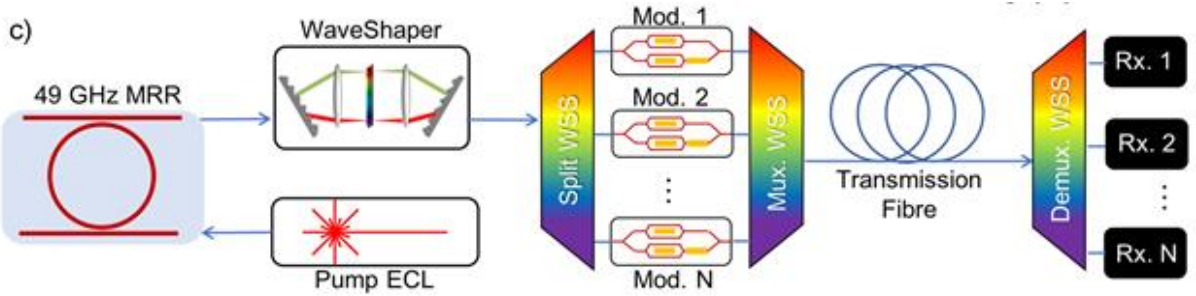

**Figure 3. Soliton crystal micro-comb communications experiment.** A CW laser, amplified to 1.8W, pumped a 48.9 GHz FSR micro-ring resonator, producing a micro-comb from a soliton crystal oscillation state. The comb was flattened and optically demultiplexed to allow for modulation, and the resulting data optically multiplexed before the subsequent transmission through fibres with EDFA amplification. At the receiver, each channel was optically demultiplexed before reception. ECL, edge-coupled laser, WSS wavelength-selective switch. Rx receiver.

Kerr optical microcombs have attracted a great deal of interest and one of their main applications has been in this area. They have successfully been used as optical sources for ultra-high bandwidth optical fiber transmission of data [24 - 26]. A key factor has been achieving the capacity to modelock all of the microcomb lines, and this has been characterized by the discovery of new states of temporal optical soliton oscillation that include feedback-stabilized Kerr combs [25], dark solitons [32] and dissipative Kerr solitons (DKS) [24]. The last one (DKS) has achieved the greatest success, being the basis of extremely high data transmission rates across the full C and L telecom bands, at a rate of 30 Tb/s using only a single source, and 55 Tb/s by using two microcombs [24].

Despite this success, though, micro-combs still need to be even more stable and simpler and robust in both operation and generation, in order to meet the demands of real-world installed fibreoptic systems [26. 28 - 32]. They particularly must work without the need for complicated stabilization feedback, preferably in uncomplicated open-loop fashion and without the need for complicated pumping schemes that DKS states need in order to be generated. Furthermore, the conversion efficiency from pump to comb lines must be much higher and their threshold pump power much lower. Systems that use microcombs also must achieve a much higher spectral efficiency (SE) since to date they have only achieved about ¼ of the theoretical maximum. Spectral efficiency is an absolutely key and fundamental parameter that limits the total data capacity of systems [28, 29].

In this paper, [26] we demonstrate a world-record high bandwidth for optical fibre data transmission using standard single mode fibre together with a single optical source. Our use of a new and powerful type of micro-comb that operates through states that have been called "soliton crystals" [11, 26], based on CMOS - compatible chips [2, 3, 33 - 50], enabled us to reach a transmission data rate of 44.2 Terabits per second using only a single chip - almost 50% greater than previously achieved [24, 25]. More importantly, we report a significant improvement by a factor of 3.7 times in the enormously important SE, achieving 10.4 bits / s / Hz which is a record high value for microcombs. We do this through the use of a very high coherent modulation format of 64 QAM, together with a microcomb that has a record low spacing, or FSR, at 48.9 GHz. We only use the telecom C-band, leaving room for significant expansion in our capacity. We report experiments in the lab with 75 km of fibre as well as over an installed optical fibre network in the greater Melbourne metropolitan area. These results were made possible because of the highly and stable and robust generation and operation of the soliton crystals, together with their very high natural efficiency. All of these features are intimately tied to the CMOS compatible nature of the integrated platform.

tightly packed self-localized pulses within micro-ring resonators [11]. Soliton crystals can occur in integrated ring resonators that have a higher order mode crossing. Further, they do not need the dynamic and very complicated pumping schemes or elaborate stabilization that self-localised DKS states need [51]. The basis of their stable behaviour originates from the fact that their intra-cavity power is dramatically higher than DKS states. In fact, it is very similar to the power levels of the chaotic temporal states [11, 52]. As a result, there is a very small difference in power levels in the cavity when the soliton crystal states are created out of chaos, and so there is no change in the resonant frequency. It is this self-induced frequency detuning arising from thermal instability due to the soliton step that renders pumping of DKS states, for example, very challenging [53]. The combined effect of natural stability and robust and simple manual generation and the overall efficiency of soliton crystals that makes them extremely attractive for very high bandwidth data transmission exceeding a Terabit per second.

## II. Experiment

A map of the metropolitan network used for the system field trial is given in Fig. 1, while the soliton crystal comb spectrum is shown in Fig. 2 and the experimental setup for the demonstration of high capacity optical data transmission in Fig. 3. The microcomb featured a 48.9 GHz FSR, producing a soliton crystal output with a spectrum spanning across > 80 nm while pumping at 1.8 watts of CW power at a wavelength of 1550nm. The soliton crystal micro-comb was preceded first by the primary comb and displayed very variation in comb line powers at < +/- 0.9 dB, for ten different incidents of initiation, and was achieved by sweeping the wavelength manually from 1550.300 - 1550.527 nm. This clearly proves the micro-comb turn-key generation repeatability for our devices.

Out of the total number of generated comb lines, eighty were chosen from the 3.95 THz, 32 nm wide C-band window at 1536 – 1567 nm. The spectrum was then flattened using a WaveShaper. Following this the number of wavelengths was doubled to 160, corresponding to a 24.5 GHz spacing, to increase the spectral efficiency. This was accomplished with a single sideband modulation technique that generated both even and odd channels that were not correlated. We then grouped six wavelengths, with the rest of the bands supporting data loaded channels based on the same even-odd structure. We were able to use a record high order 64 QAM coherent modulation format that modulated the whole comb at a baud rate of 23 Giga-baud, that achieved 94% utilization of the available spectrum.

We performed 2 experiments, the first across 75 km of single mode optical fiber in the lab and the second in a field trial using a metropolitan network in the greater Melbourne area, also based o standard SMF (Fig. 1), which linked Monash University's Clayton campus to the RMIT campus in the Melbourne CBD. The signal was recovered at the receiver with a standard offline digital signal processor (DSP). The constellation diagrams (Fig. 4) at 194.34 THz for the back-to-back configuration where the transmitter was connected directly to the receiver, show that the quality of the signal as reflected in Q2, from error vector magnitude, was almost 18.5 dB, decreasing slightly to 17.5 dB the entire set of comb lines were modulated across the span.

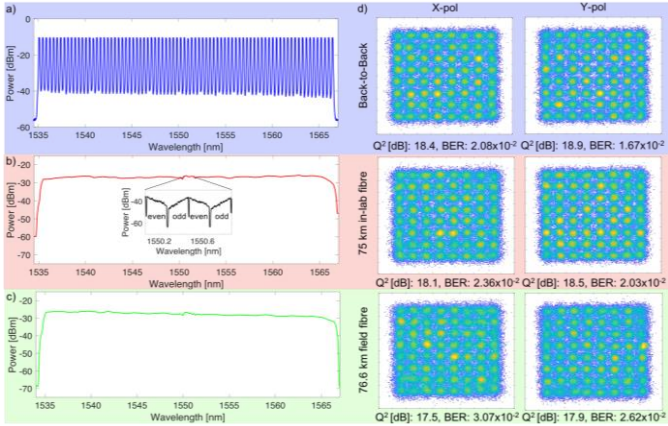

**Figure 4.** a-c) Spectra of the soliton crystal frequency comb after flattening (a), modulation and transmission through either 75 km spooled in-lab fibre (b) or through the field-trial link (c). The spectrum (a) is measured with 12.5 GHz resolution to resolve the individual comb lines, while (b) and (c) are plotted at 50 GHz resolution to illustrate average channel powers. Flattening equalised the comb line power to within 1 dB. After modulation and amplification, the channels were shaped by the EDFA gain spectrum. The inset in (b) depicts the test channel spectra captured with a 150 MHz resolution optical spectrum analyser, highlighting the odd and even sub-bands modulated onto each comb line in the test band. d) Constellation diagrams for a comb line at 193.4 THz (1550.1 nm) for both X- and Y-polarization channels. 'Back to back' denotes the transmitter directly connected to the receiver, '75 km in-lab fibre' indicates reception after transmission through 75 km of spooled fibre inside the lab, while '76.6 km field fibre' denotes reception after transmission through the field-trial link. BER and $Q^2$ related to the constellations are noted.

### III. RESULTS AND DISCUSSION

The performance of the transmission as measured by the bit error ratio (BER) as a metric for each channel is shown in Figure 5. We studied 3 cases: i) directly connecting the receiver to the transmitter stages termed back-to-back (B2B), following transmission across the ii) fiber in the lab and iii) transmitting the data across the installed metro area network. The performance for all the channels was degraded by transmission, but this was anticipated. Figure 5a shows the 20% threshold for soft-decision forward error correction (SD-FEC), a common benchmark for performance, using a proven code, at a BER of $4\times10^{-2}$ [53]. All measurements achieved under the FEC limit. However, since SD-FEC thresholds based on BER can be less accurate at higher modulation formats as well as at higher BERs [54-55], we also used generalized mutual information (GMI) to determine the performance of the system. Figure 5 shows the GMI for every channel as well as its corresponding SE, where we include lines to indicate the projected overheads. We succeeded in demonstrating a line bit rate (raw bitrate) of 44.2 Terabits per second, which corresponds to a net coded rate at 40.1 Terabits/s (for B2B), which dips to 39.2 and 39.0 Terabits/s in the lab and metro network trials, respectively. We also achieved SEs reaching 10.4, 10.2 and 10.1 bits / s / Hz.

Our data rate is an increase of almost 50% compared to the highest previously reported values achieved with a single source [26]. Even more importantly the SE is enhanced even more, being a factor of 3.7 higher than previous reports. This is quite extraordinary since we conducted the experiments under the most challenging of conditions. This includes the absence of any closed loop feedback systems or external stabilization, as well as without the use of any complex generation pump methods. On top of this, we actually fully flattened or equalized, the comb lines, even though this was not required [56]. We did this because we wanted to address any possibilities of the non-flat soliton crystal comb spectrum being construed as representing any sort of limitation. Since we performed our experiments with the use of comb flattening, and this was not necessary, then doing the experiments without flattening would only reduce the system impairments and would actually improve our results even further. Hence, in doing this we clearly show that having a nonuniform spectrum is not any sort of limitation. This identical line of argument applies to the issue of not needing any closed-loop feedback control for the micro-comb. We could always include this as all other experiments have done, and this would again improve our performance even more.

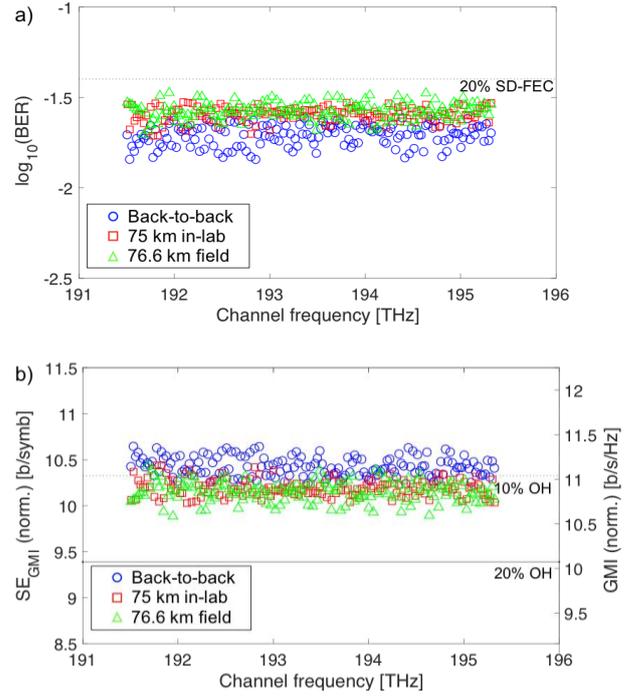

**Figure 5.** a) BER for each comb line. Blue circles points indicate performance of channels in B2B configuration, red squares dots are for performance after transmission through 75 km of in-lab spooled fibre, while green triangles are after transmission through the 76.6 km installed metropolitan-area fibre link. An indicative FEC threshold is given at $4\times10^{-2}$, corresponding to a pre-FEC error rate for a 20% soft-decision FEC based on spatially-coupled LDPC codes [25] (dashed line). After transmission, all channels were considered to be error-free, b) GMI and spectral efficiency measured for each comb line. GMI was calculated after normalization to scale measured constellations in order to account for received signal-to-noise ratio (SNR). Lines are for 20% and 10% overheads. Spectral efficiency was derived from GMI, and the ratio of symbol rate to comb spacing. GMI indicates a higher overall capacity than BER with the indicated SD-FEC threshold, as GMI assumes the adoption of an ideal code for the system. For B2B, GMI (SE) varied between 11.3 b/symb. (10.6 b/s/Hz) and 10.9 b/symb. (10.3 b/s/Hz). After in-lab fibre transmission, the achievable per-channel GMI (SE) varied between 11.0 b/symb. (10.4 b/s/Hz) and 10.7 b/symb. (10.1 b/s/Hz), with the same range observed for the installed field-trial fibres. We estimate the overall capacity from the sum of the GMIs, multiplied by the symbol rate.

The record high spectral efficiency and absolute bandwidth that we achieve were greatly aided by the very high conversion efficiency we achieved between the pump and the soliton crystal comb lines [11, 52]. Again, as mentioned this results from the very small power step in the cavity that occurs when the soliton crystals are generated from the chaos states.

We only used the telecom C-band, and yet the bandwidth of the microcomb was larger than 80 nm. Therefore, wavelengths in both the L (1565-1605 nm) and even S (1500-1535 nm) bands could easily be used. In fact even broader bandwidths can be achieved by increasing the power, by varying the wavelength of the pump, by engineering the dispersion or further methods. This would yield an increase of more than a factor of 3 in total bandwidth, resulting in >120 Terabits per second using only a single source.

Achieving even lower spacings, or FSRs, with miro-combs would yield yet higher SEs since the quality of the signal increases for smaller baud rates. This may result in a smaller overall comb bandwidth however. For our experiments, the use of single sideband modulation allowed multiplexing two channels using one single wavelength, which cut the comb spacing by a factor of 2 while enhancing the back-to-back performance that was limited by transceiver noise. This was made possible by the stability of the soliton crystals. Conversely, electro-optic modulation has also been used to sub-divide the micro-comb repetition rate, and this would also create broader comb bandwidths. This, however, would require locking the comb FSR spacing to an external RF source, although this is feasible since sub megahertz stabilization of microcombs has been achieved [57, 58]. Furthermore, increasing the comb conversion efficiency by using a newly discovered class of soliton, called laser cavity-soliton micro-combs [34] will offer a powerful way to increase the system capacity as well as the quality of the signal even further. For recently installed networks, our approach can easily be complemented by using spatial division multiplexing based on multiple core fibre [29, 59], yielding bandwidths of more than a petabit per second using a single microcomb. Our results join the many breakthroughs achieved with microcombs, and in particular using soliton crystal combs. These particularly include our applications of soliton crystals to RF and microwave signal processing [60 - 81]. This work presented here is the most challenging demonstration ever reported for micro-combs in terms of ease of generation, coherence, stability, noise, efficiency, and others, and is a direct result of the superior soliton crystal microcomb qualities.

## IV. Conclusions

We demonstrate a new world record for performance of ultra-high bandwidth optical transmission systems using a single optical source over standard optical fiber. We achieve this through the use of soliton crystal micro-combs that have a very low FSR spacing of 48.9GHz. Our achievement results from this record low comb spacing together with the efficient, broad bandwidth, and stable nature of soliton crystals, together with their CMOS compatible integration platform. Soliton crystal micro-combs are fundamentally low noise and coherent and can easily be initialised and operated using only very simple open-loop control that only requires commercially available components. Our results clearly show the ability of soliton crystal microcombs to achieve world record high bandwidths for optical data transmission over fibre in very demanding real-world applications.